# Silicon-vacancy color centers in phosphorus-doped diamond

Assegid Mengistu Flatae[1*], Stefano Lagomarsino[1,2], Florian Sledz[1], Navid Soltani[1], Shannon S. Nicley[3,4], Ken Haenen[3], Robert Rechenberg[5], Michael F. Becker[5], Silvio Sciortino[2,6], Nicla Gelli[2], Lorenzo Giuntini[2,6], Francesco Taccetti[2] and Mario Agio[1,7]

[1] *Laboratory of Nano-Optics and Cμ, University of Siegen, 57072 Siegen, Germany*
[2] *Istituto Nazionale di Fisica Nucleare, Sezione di Firenze, 50019 Sesto Fiorentino, Italy*
[3] *Institute for Materials Research (IMO) & IMOMEC, Hasselt University & IMEC vzw, 3590 Diepenbeek, Belgium*
[4] *Department of Materials, University of Oxford, Parks Road, Oxford, OX1 2PH, United Kingdom*
[5] *Fraunhofer USA Center for Coatings and Diamond Technologies, East Lansing, MI 48824, USA*
[6] *Dipartimento di Fisica e Astronomia, University of Florence, 50019 Sesto Fiorentino, Italy*
[7] *National Institute of Optics (INO), National Research Council (CNR), 50125 Florence, Italy*

\* flatae@physik.uni-siegen.de

**Abstract:** The controlled creation of color centers in phosphorus-doped (n-type) diamond can facilitate the electronics integration of quantum photonics devices, such as single-photon sources operating upon electrical injection. Silicon vacancy (SiV) color centers are promising candidates, but so far the conditions for single-photon emission in phosphorus-doped diamond have not been investigated. In this study, we create SiV color centers in diamond samples with different phosphorus concentrations and show that the fluorescence background due to doping, nitrogen-impurities and ion implantation induced defects can be significantly suppressed. Single-photon emitters in phosphorus-doped diamond are obtained at the low Si-ion implantation fluences.

1. Introduction

Efficient and scalable solid-state single-photon sources operating at room temperature are crucial to the development of quantum photonics technologies, such as quantum computers, secure communication lines and precision measurements below the shot-noise limit [1-3]. In these respects, optically and electrically driven semiconductor quantum dots exhibit the best performances, but they require operation at cryogenic temperatures [4-7]. Under ambient conditions, color centers in diamond may represent an alternative platform. In particular, silicon vacancy (SiV) color centers have most of the fluorescence emission (>90%) concentrated in a narrow zero-phonon line at room temperature [8]. In addition, the center exhibits a short excited-state lifetime and a small inhomogeneous linewidth broadening [8].

It has been recently shown that the electrical excitation of SiV color centers is determined by electron and hole-capture processes, which are directly related to the density of free electrons and holes in the vicinity of the color center [9]. This possibility has been recently demonstrated using p-i-n based diode structures [10,11], where the Si-ions are implanted in the intrinsic layer. However, the creation of SiV color centers in p- or n-type diamond can avoid the need for the complex semiconductor junctions. For example, SiV color centers in n-type diamond allows the implementation of a Schottky-diode. This would give the possibility of injecting holes directly from the metal contact and of utilizing electrons from the n-type diamond host lattice. In recent years, n-type conduction in diamond has been successfully obtained in phosphorous (P)-doped diamond grown by microwave plasma chemical vapor deposition (CVD) [12-15].

However, there are challenges to attain the conditions for single-photon emission in P-doped diamond. The crystal imperfections (structural distortion and degradation caused by the incorporation of phosphorus atoms during the growth process [16-21]), phosphorus-based bound excitons [22] and phosphorus-related complex defects [23,24], photoluminescence related to phosphorus-nitrogen co-doping [23,25] and ion-beam induced defects can compromise the signal-to-noise ratio required for single-photon emission.

Previous photoluminescence experiments on P-doped diamond revealed broad background emission assigned to donor-acceptor-pair recombination although the nature of the defects was not clear. CVD diamond crystal usually contains a high concentration of vacancies, as high as $10^{18}$ cm$^{-3}$ [26], and some form of phosphorus and vacancy related bound up is a possibility. The lattice damage caused during ion implantation can be partially healed by annealing. Its effectiveness in favoring Frenkel couples recombination that avoid the formation of vacancy complexes have been reported for p-type (boron-doped) diamond [27], but reproducible results in P-doped diamond have not yet been reported [28]. Therefore, attaining single-photon emission from SiV color centers in P-doped diamond is not obvious.

Here, we investigate the optical properties of SiV color centers in P-doped single-crystal diamond under different implantation and doping parameters, namely Si-ions implanted at a fluence range of $10^7$-$10^{14}$ cm$^{-2}$ and phosphorus concentrations in the gas feed of 4300 ppm, 5000 ppm and 20000 ppm. We show that the luminescence background can be significantly suppressed even for high phosphorus doses and that the conditions for single-photon emission can be attained.

2. Samples

Three P-doped single-crystal diamond films were grown on different high-pressure-high-temperature (HPHT) diamond substrates by microwave plasma-enhanced CVD (MWPECVD) using phosphine (PH$_3$) as a dopant source.

One of the P-doped singe crystalline diamond thin film was grown on a (111) oriented HPHT diamond substrate with lateral dimensions of 2.6 mm x 2.6 mm (sample A). Prior to deposition, the growth surface was mechanically polished using a high speed scaife. The P-doped diamond was grown utilizing an in-house built 2.45 GHz MWPECVD reactor at a pressure of 160 Torr and 2 kW absorbed microwave power. The resulting deposition temperature was 940 °C, as measured by a single color IR-pyrometer (ε = 0.6). The hydrogen (H$_2$) rich plasma contained 0.09 % of methane (CH$_4$) with a PH$_3$/CH$_4$ ratio of 4300 ppm. 4 hrs of deposition resulted in a smooth diamond film with a thickness of 2.0 µm +/- 0.1 µm measured by linear encoder (Solartron DR600).

The other two P-doped homo epitaxial samples were grown similarly on (111) oriented HPHT diamond substrates with lateral dimensions of 2.5 mm x 2.5 mm (Sumitomo). Both samples were grown by MWPECVD in a 2.45 GHz ASTeX PDS17 reactor with a water cooled molybdenum substrate holder. A 500 sccm total flow rate and pressure of 140 Torr were held constant. The H$_2$ plasma contained 0.15%

CH$_4$. One of the samples (sample B) was grown with a P-gradient, with the PH$_3$/CH$_4$ ratio adjusted in eight one hour-steps, of 0, 250, 500, 1000, 2500, 5000, 10000, and 20000 ppm, respectively. The other sample (sample C) was grown with a constant 5000 ppm PH$_3$/CH$_4$. The applied forward microwave power was varied between 1.0 and 1.3 kW to maintain a substrate growth temperature of 1000 °C. The substrate temperature was also monitored by tungsten vanishing filament pyrometry assuming an emission coefficient of $\epsilon$ = 0.6. Both samples were grown for 8 hours of total deposition time. The H$_2$ and CH$_4$ gasses are filtered to < 1 ppb (9 N) purity, and the PH$_3$ is used from a source diluted to 200 ppm in H$_2$. The sample morphologies of the films were investigated by optical microscopy and the thicknesses were measured by linear encoder (Mitutoyo 542-158), with nine measurements at the sample center, with the thickness error estimated as the average of the standard errors before and after growth. Sample B was grown to a thickness of 1.3 µm ± 0.2 µm and sample C to 2.2 ± 0.2 µm.

The three P-doped samples were used as a host matrix for the creation of SiV color centers. The Si-ions implantation is based on a 3 MeV Tandetron accelerator (High Voltage Engineering Europe) equipped with a HVEE 860 Negative Sputter Ion Source. This allows for the acceleration of ion species (Si$^+$, Si$^{2+}$, Si$^{3+}$) in the range of few MeV energies depending on the terminal voltage and the ion charge state. Aluminum (Al) metal foils were used to decrease the ion energy down to a few tens of keV for shallow implantation (within 200 nm from the surface). The possibility of using pulsed ion-beams allow control over the number of implanted ions in the desired place. A custom designed furnace (1200 °C in high-vacuum conditions ~ 10$^{-7}$ mbar) enables the activation of the SiV color centers in P-doped samples. The samples have been Si-implanted with five different fluences (~10$^{14}$ cm$^{-2}$, ~10$^{13}$ cm$^{-2}$, ~10$^{12}$ cm$^{-2}$, ~10$^{8}$ cm$^{-2}$ and ~10$^{7}$ cm$^{-2}$). The expected implantation depth is ≤ 200 nm. Details on the implantation process can be found in Ref. [29]. To optically characterize the samples, spectrally and temporally resolved µ-photoluminescence measurements are conducted in a homemade confocal microscopy setup equipped with a Hanbury-Brown Twiss interferometer for the verification of single-photon emission [8].

3. Experimental results

3.1 Spectroscopy and discussion

Sample A was implanted with four different Si-ion fluences of 10$^8$ cm$^{-2}$, 10$^{12}$ cm$^{-2}$, 10$^{13}$ cm$^{-2}$ and 10$^{14}$ cm$^{-2}$. Unlike for intrinsic single-crystal diamond [8], fluence-dependent background at the emission wavelength of the SiV color centers is evident, especially for the higher implantation fluences, as depicted in Fig. 1a. The peak at 708 nm corresponds to the Raman line of diamond excited by a 647-nm laser. Using two band pass filters, the SiV color centers's emission spectral window is selected and the decay kinetics are measured. The fluorescence lifetime shows a background-related slow component (8.0-8.6 ns) and a fast component corresponding to the excited-state lifetime of the SiV (0.6-1.3 ns). Figure 1b shows the measurement in different parts of the sample. Depending on the implantation fluence and on the observed region (the ion distribution for a given fluence is not uniform [8,29]), the contribution from the background changes. The region implanted with 10$^8$ cm$^{-2}$ shows relatively low background, comparable to the region, where no ions are implanted. Figure 1c shows the emission from the color centers (black curve) and the background (bg) far away from the implantation spot (blue curve).

Unlike for higher implantation fluences, the fluorescence decay kinetics in the region implanted with $10^8$ $cm^{-2}$ exhibits a single exponential decay with a time constant of 1.0 ns, corresponding to the excited-state lifetime of the SiV in intrinsic single-crystal diamond [8]. Changes in the local environment can be the reason for observing a slight fluctuation in the fluorescence lifetime of the SiV color centers in P-doped diamond. In the region with the lowest implantation fluence, the emitters are well separated as shown in the wide field image of Fig. 1d inset. The emitters, however, do not exhibit anti-bunching behavior, indicating a possible clustering of two or more centers. With such implantation fluence, the expected ions density is ~ 1 $\mu m^{-2}$. Taking into consideration a few percent activation yield (1-3% [8]), clustering of SiV centers is evident. Color center clustering is common at the grain boundaries of polycrystalline diamond, but in a P-doped diamond crystal the unavoidable presence of defects (dislocations) can be a possibility [30]. We have also observed the analogous phenomenon for SiV color centers in intrinsic single-crystalline diamond [8, 29].

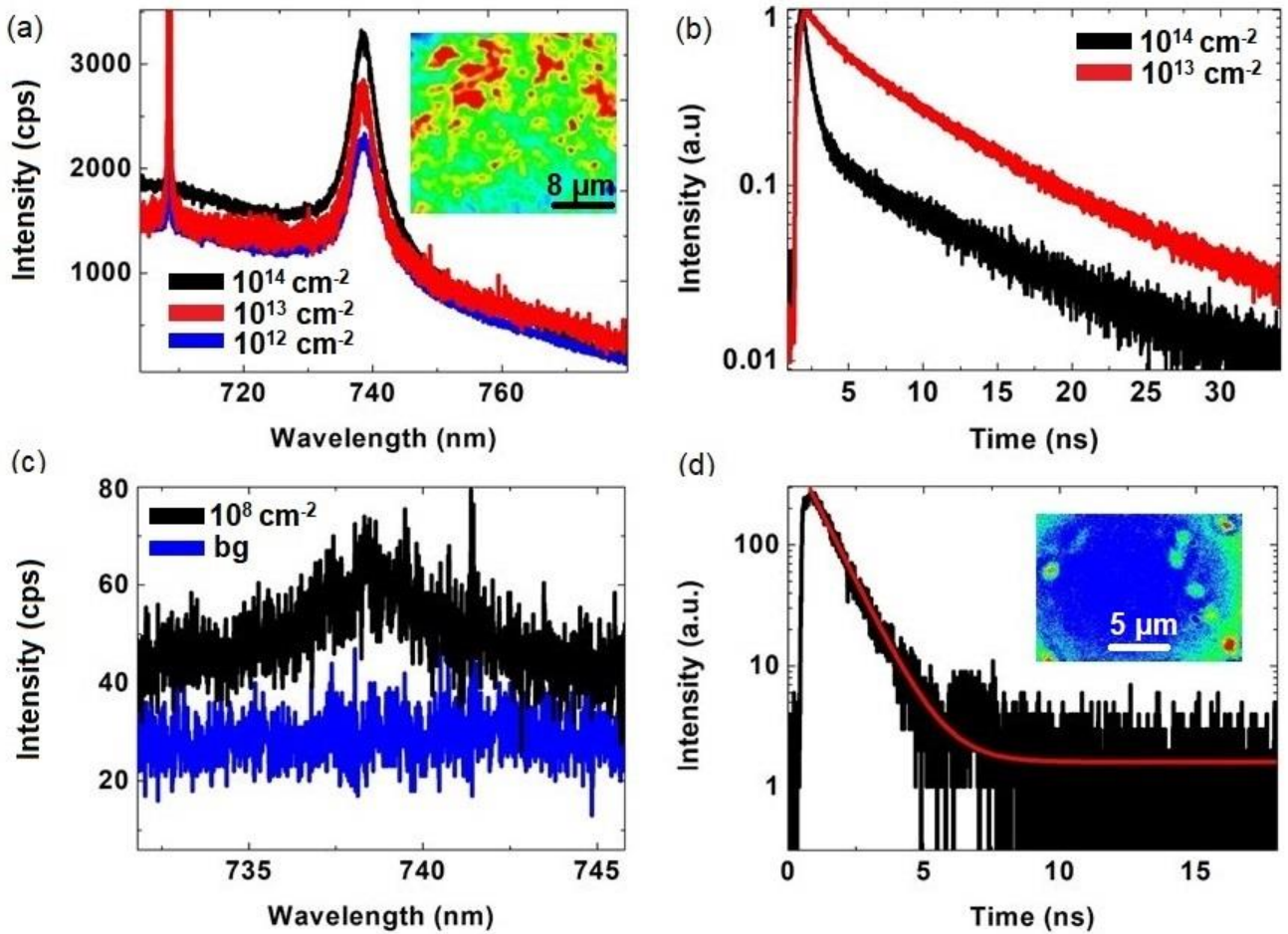

Figure 1. (a) The photoluminescence spectra of SiV centers in the region of sample A implanted with a fluence of $10^{12}$ $cm^{-2}$, $10^{13}$ $cm^{-2}$ and $10^{14}$ $cm^{-2}$. Inset: wide-field imaging of SiV color centers implanted with $10^{14}$ $cm^{-2}$. (b) The fluorescence lifetime measurement exhibits a bi-exponential decay. The short lifetime corresponds to SiV color centers and the long lifetime corresponds to temporally-correlated

background. (c) The photoluminescence spectrum of SiV centers implanted with a fluence of $10^8$ cm$^{-2}$ (black curve) and background (bg) away from the implantation region (blue curve). (d) The fluorescence lifetime measurement in the region of sample A implanted with a fluence of $10^8$ cm$^{-2}$ exhibits single exponential decay with a time constant of 1.0 ns, corresponding to the excited-state lifetime of SiV color centers.

The ion-beam fluence dependent background is further investigated by different laser excitations (532 nm, 647 nm, 656 nm and 690 nm) and we attribute its main source to the creation of nitrogen-vacancy (NV) related impurities. A higher fluence of implanted Si-ions creates more vacancies and favors the formation of NV color centers during annealing. Figure 2a shows the photoluminescence spectra in the region implanted with a fluence of $10^{12}$ cm$^{-2}$, $10^{13}$ cm$^{-2}$ and $10^{14}$ cm$^{-2}$ using 532 nm CW laser excitation at 2 mW. The increase in the emission of NV color centers with increasing Si-ion implantation fluence is clearly observed. Spatially- and spectrally-resolved confocal images show that the NV color centers are mainly created at the location of SiV color centers, as depicted in Fig. 2b, c and d. A precise information about the position of the sample (along z-direction or its thickness) and its corresponding fluorescence signal enables to collect both spatial- and spectral-information about the SiV color centers and the background at different depths. In the figures, the vertical-axis is centered at the position, where the SiV color centers exhibit the highest fluorescence signal, whereas the positive and negative positions represent the scanning direction from air to diamond and inside diamond, respectively. The accurate determination of the implantation depth is subjected to the convolution of both the spread in the confocal volume and in the implantation depth. The background due to NV color centers is reduced by exciting the SiV color centers using 656 nm and 690 nm CW laser, as NV complexes have less absorption in this spectral range. Figure 3a shows the photoluminescence spectra of the color centers using 690 nm CW laser excitation (at 1 mW). Even if the background is reduced, its presence is still clearly observed in the spatially- and spectrally-resolved images (Fig. 3b, c, d). The line around 760 nm is the Raman signal of diamond (1332 cm$^{-1}$) corresponding to the 690 nm laser excitation.

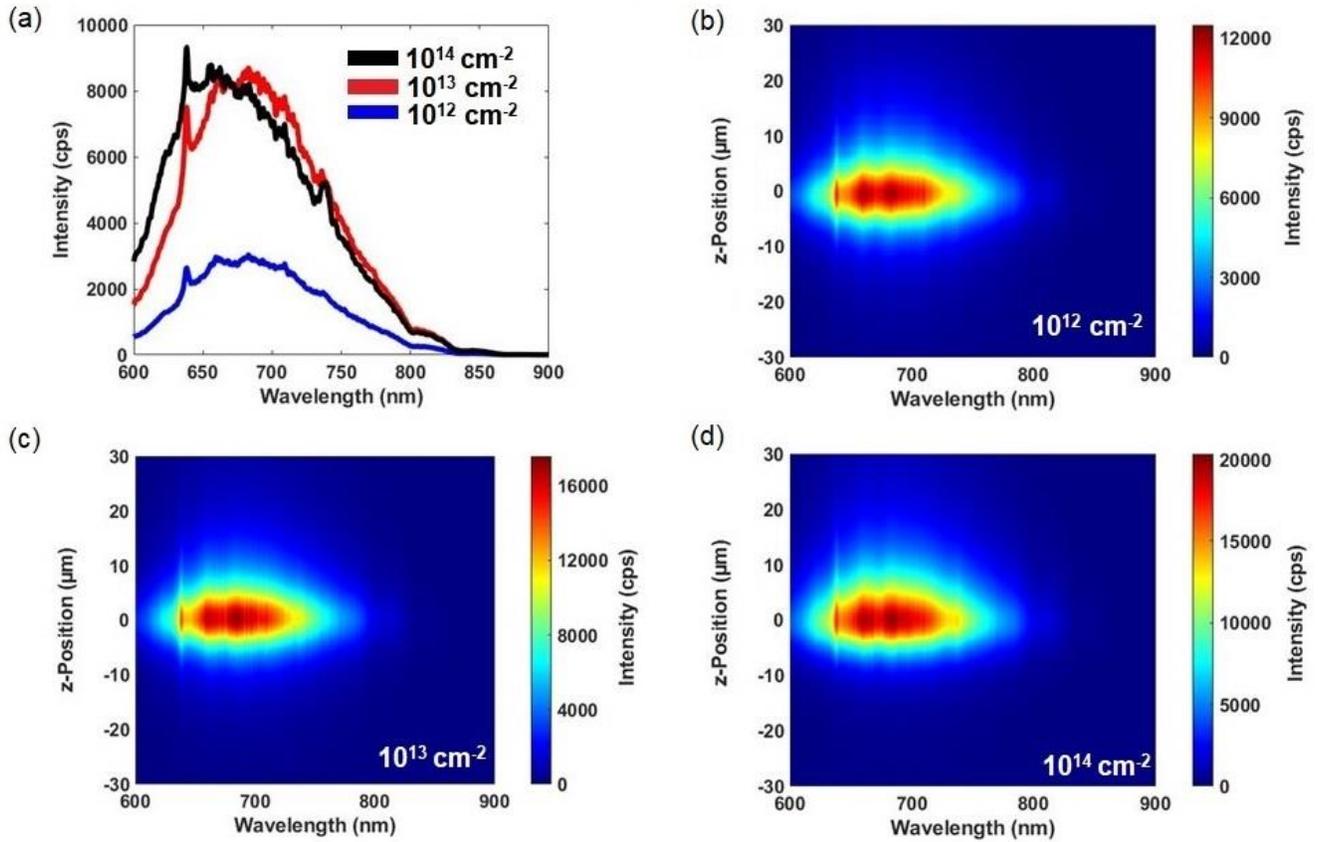

Figure 2. (a) The photoluminescence spectra in the regions of sample A implanted with a fluence of $10^{12}$ cm$^{-2}$, $10^{13}$ cm$^{-2}$ and $10^{14}$ cm$^{-2}$ using 532 nm CW laser show NV-related background. (b), (c), (d) Spatially- and spectrally-resolved confocal images indicate that the NV color centers are mainly created at the location of the ion damage and that they are dependent on the implantation fluence.

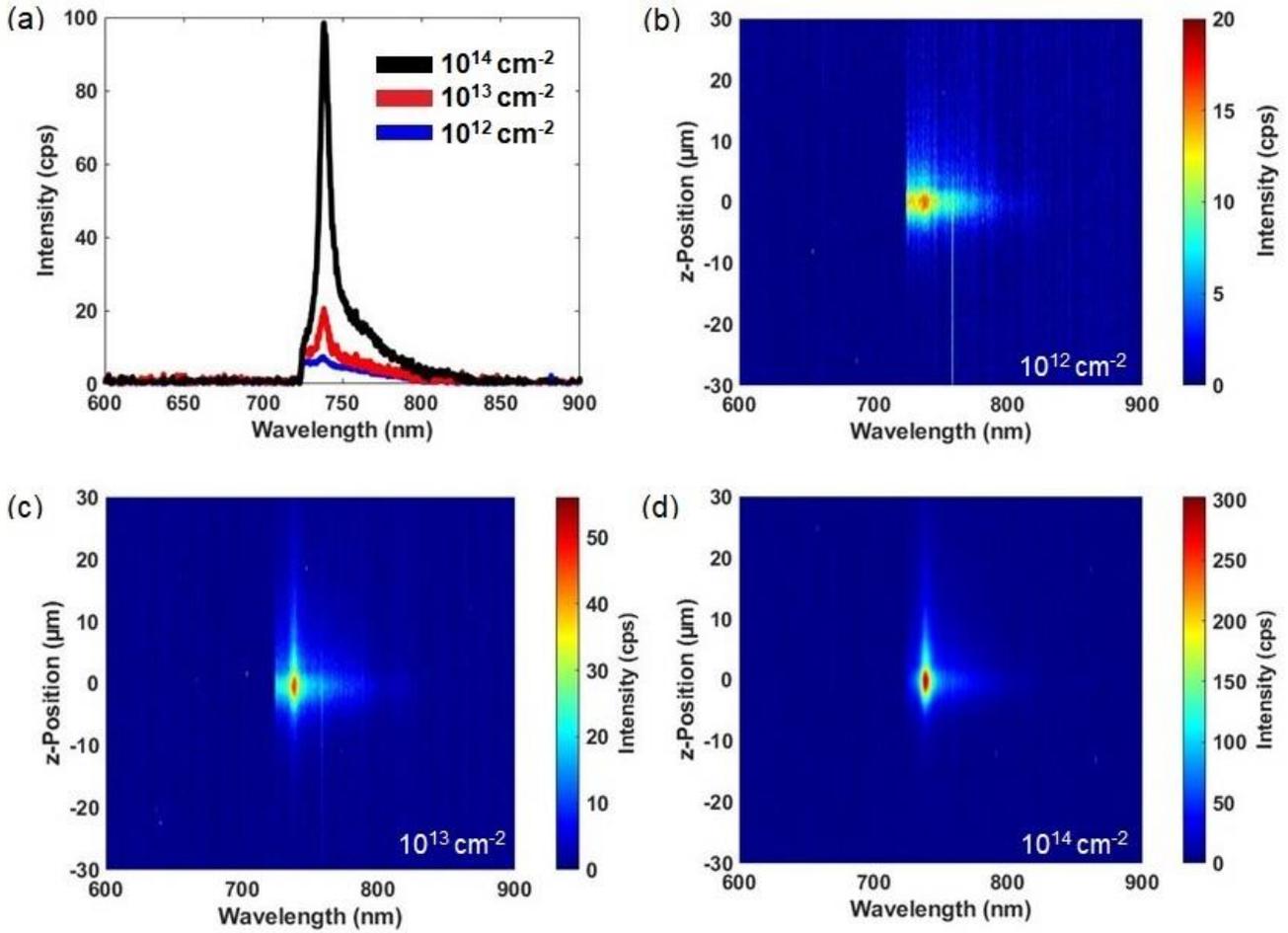

Figure 3. (a) The photoluminescence spectra in the regions of sample A implanted with a fluence of $10^{12}$ cm$^{-2}$, $10^{13}$ cm$^{-2}$ and $10^{14}$ cm$^{-2}$ using 690 nm CW laser show NV-related background. (b), (c), (d) Spatially- and spectrally-resolved images indicate that the NV color centers are mainly created at the location of defect formation and that they are dependent on the implantation fluence. The NV contribution is reduced due to the smaller absorption coefficient of NV centers at 690 nm. The line around 760 nm is the Raman signal of diamond (1332 cm$^{-1}$) corresponding to the 690 nm laser excitation.

Similar implantation dependent background is also observed for sample B. This sample was implanted with Si-ion fluences of ~$10^{12}$ cm$^{-2}$ and ~$10^{7}$ cm$^{-2}$. Even if the background is not as pronounced as in the previous sample (sample A), the photoluminescence spectrum and the fluorescence lifetime measurements confirm the trend. Figure 4a depicts the photoluminescence spectra of the color centers in different parts of the sample. Due to the Gaussian-like profile of the implanted ion distribution, the color centers in the tail of the $10^{7}$cm$^{-2}$ region exhibit less background and have similar optical properties as for intrinsic single-crystal diamond (blue curve) [8]. In the highly implanted region (Fig. 4a, black curve) the background is still observable and its temporal contribution is present in the fluorescence decay curve as shown in Fig.

4b. The slow component is attributed to the background and the fast one to the excited-state lifetime of the SiV (0.8 ns).

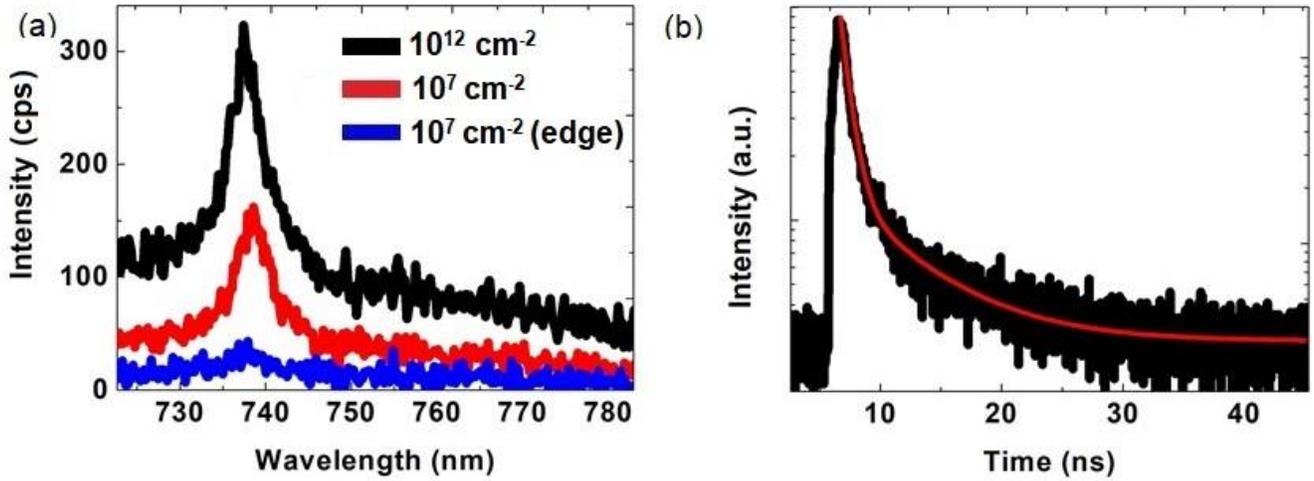

Figure 4. (a) The photoluminescence spectra of SiV centers in sample B implanted with a fluence of $10^7$ cm$^{-2}$ and $10^{12}$ cm$^{-2}$. (b) The fluorescence-lifetime measurement of the region implanted with $10^{12}$ cm$^{-2}$ exhibits a bi-exponential decay. The short lifetime (0.8 ns) corresponds to SiV color centers and the long one to temporally-correlated background.

For further investigation of SiV color centers in P-doped diamond and for attaining single-photon emission, sample C was implanted with Si-ions at four different fluences (~$10^{14}$ cm$^{-2}$, ~$10^{13}$ cm$^{-2}$, ~$10^{12}$ cm$^{-2}$ and ~$10^7$ cm$^{-2}$). Spectral measurements show that the implanted regions exhibit less background (Fig. 5a) and, where the implanted ions are fewer ($10^7$ cm$^{-2}$ and in the tail of a region implanted with $10^{12}$ cm$^{-2}$), the signal is dominated by SiV emission. The inset in Fig. 5a shows the emission of the SiV color centers (blue curve) and the background (bg) away from the implanted region (black curve). The background in this region is not temporally-correlated and leads a mono-exponential fluorescence decay kinetics, which corresponds to the excited-state lifetime of SiV color centers (1.3 ns), as shown in Fig. 5b. This sample is further investigated for single-photon emission as discussed in the next session.

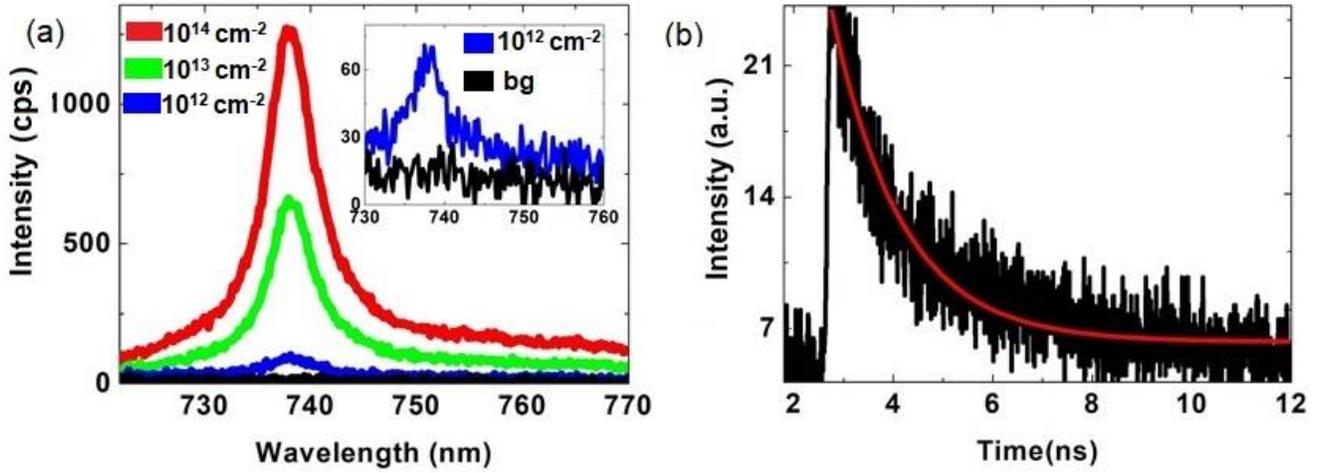

Figure 5. (a) The photoluminescence spectra of SiV centers implanted with a fluence of $\sim 10^{12}$ cm$^{-2}$, $\sim 10^{13}$ cm$^{-2}$ and $\sim 10^{14}$ cm$^{-2}$. Inset: Spectrum of SiV centers implanted with a fluence of $\sim 10^{12}$ cm$^{-2}$ (blue curve) and the background (bg) signal away from the implantation regions of sample C (black curve). (b) The fluorescence lifetime measurement of the region implanted with low fluence exhibits single exponential decay with a time constant of 1.3 ns, which corresponds to the excited state lifetime of SiV color centers.

The decrease in background is mainly attributed to the creation of less NV color centers. Further investigation of the sample using 532 nm, 647 nm, 656 nm and 690 nm laser confirms our claim. Figure 6a shows the photoluminescence spectra in the regions of sample C implanted with a fluence of $10^{12}$ cm$^{-2}$, $10^{13}$ cm$^{-2}$ and $10^{14}$ cm$^{-2}$ using 532 nm CW laser excitation at 2 mW. The increase in the contribution of NV color centers with increasing in Si-ion implantation fluence is clearly observed, but less pronounced as compared to sample A. Spatially- and spectrally-resolved confocal images also show a similar trend at the location of SiV color centers, as depicted in Fig. 6b, c and d. The background due to NV color centers was further reduced by exciting the SiV color centers using 656 nm and 690 nm CW laser. Figure 7a shows the photoluminescence spectra of the color centers using 690 nm CW laser excitation at 1 mW. The decrease in the background is also clearly observed in Fig. 7b, c and d.

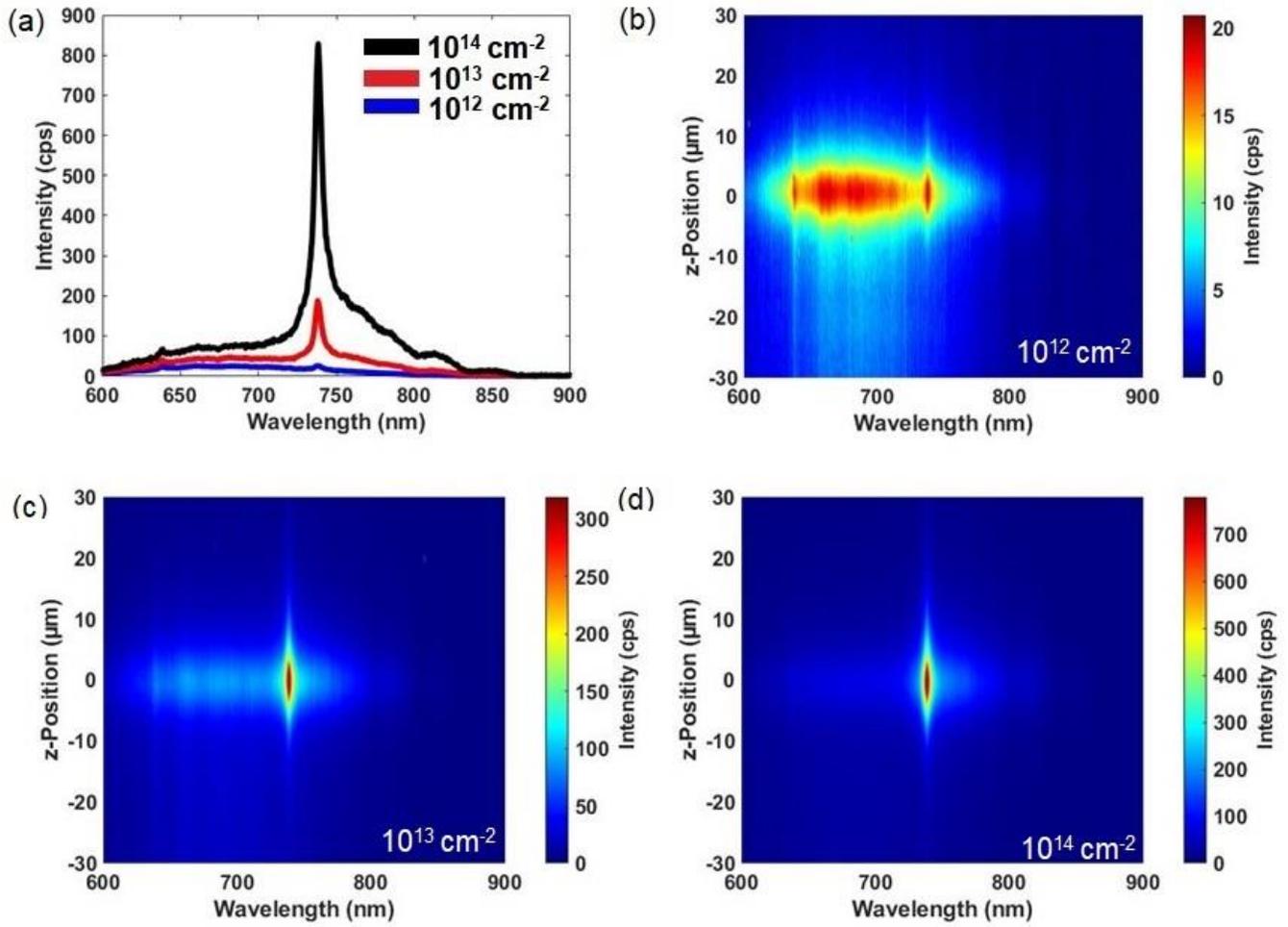

Figure 6. (a) The photoluminescence spectra in the regions of sample C implanted with a fluence of $10^{12}$ cm$^{-2}$, $10^{13}$ cm$^{-2}$ and $10^{14}$ cm$^{-2}$ using 532 nm CW laser shows NV related background. (b), (c), (d) Spatially- and spectrally-resolved confocal images show that the NV color centers are mainly created at the location of ion-induced defect formation and that they are dependent on the implantation fluence.

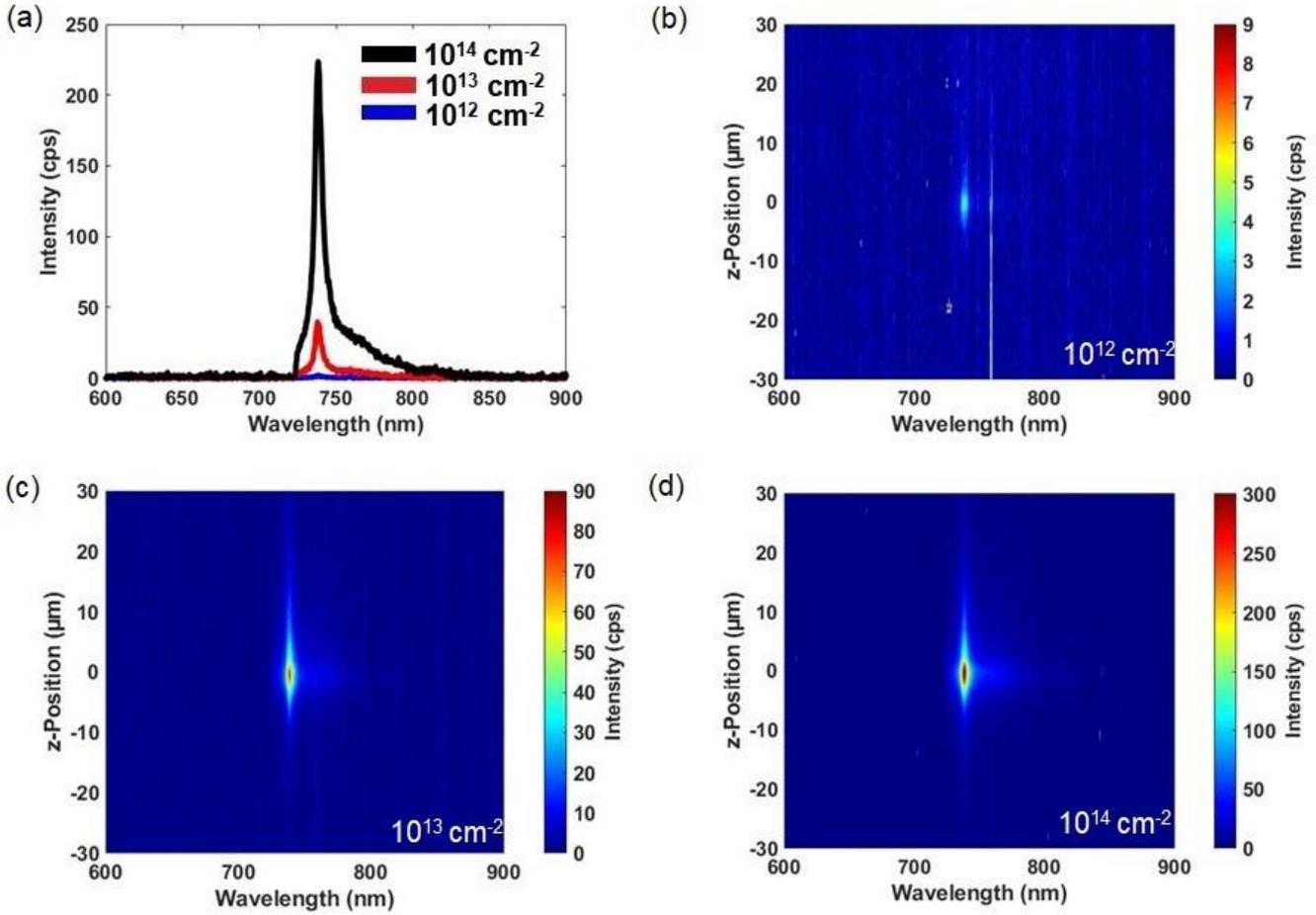

Figure 7. (a) The photoluminescence spectra in the region of sample C implanted with a fluence of $10^{12}$ cm$^{-2}$, $10^{13}$ cm$^{-2}$ and $10^{14}$ cm$^{-2}$ using 690 nm CW laser shows NV related background. (b), (c), (d) Spatially- and spectrally-resolved images show that the NV color centers are mainly created at the depths of damage formation and that they are dependent on the implantation fluence. The line around 760 nm is the Raman signal of diamond (1332 cm$^{-1}$) corresponding to the 690 nm laser excitation.

In general, as shown clearly in the temporally-, spatially- and spectrally-resolved measurements, sample C has relatively less background (supposedly due to less nitrogen content) and it is further investigated for single-photon emission as discussed in the next session.

3.2 Single-photon emission

A wide-field imaging in the tail of a region implanted with $10^{12}$ cm$^{-2}$ and in the region implanted with the lower fluence $10^7$ cm$^{-2}$ of sample C shows well separated SiV color centers, as depicted in Fig. 8a. Each of them is investigated and the signal-to-noise ratio of a SiV center is optimized for the detection of single photons. Figure 8b shows the spectrum of a single SiV color center. A Gaussian fit (red curve) yields a full-width at half-maximum (FWHM) of 7.6 nm. This SiV color center has been identified as a single-photon source by measuring the 2$^{nd}$ order intensity autocorrelation function $g^2(\tau)$ of the emitted photons in pulsed excitation (pulse width < 90 ps), as a function of the time delay τ. At the repetition rate of 20 MHz, g²(τ) exhibits individual fluorescence pulses spaced by the pulse period with a missing one at τ equal to zero, which confirms a single-photon source (Fig. 8c). Since due to electronic and photonic noises, $g^{(2)}(0)$ cannot be exactly zero, a commonly accepted signature of a single-photon source is $g^{(2)}(0) < 0.5$. Figure 8c reveals the observation of a single-photon source in a P-doped diamond sample with $g^2(0) = 0.3$. To the best of our knowledge, this is the first observation of single-photon emission from this kind of system. Most of the emitters shown in Fig. 8a do not show an anti-bunching behavior. This is attributed to the tendency of the activated centers to cluster. The presence of defects (e.g. dislocations) around which the formation of a SiV center could be energetically favored could be an explanation. These points are currently under investigation.

The intensity-dependent background of the sample prevents the observation of single-photon emission at high excitation power (above saturation). The red curve in Fig. 8d shows the count rate of the SiV color center as a function of the excitation power. The count rates are background corrected (including the detector's dark counts). The fluorescence count rate I of an emitter as a function of the excitation power $P$ is described as $I = I_\infty/(1+P_{sat}/P) + C_{bg}P$, where $P_{sat}$ is the excitation power at saturation and $I_\infty$ is the maximum photon count rate. The background $C_{bg}P$ is measured in the nearby region, where there is no SiV color center (black curve). Using this equation, we found a maximum photon count rate of $I_\infty = 1.576 \pm 0.035 \times 10^3$ cps (at $P_{sat} = 36$ μW $\pm$ 3 μW), which corresponds to about ~ $4 \times 10^5$ cps taking into account light trapping in diamond and the overall detection efficiency of our setup.

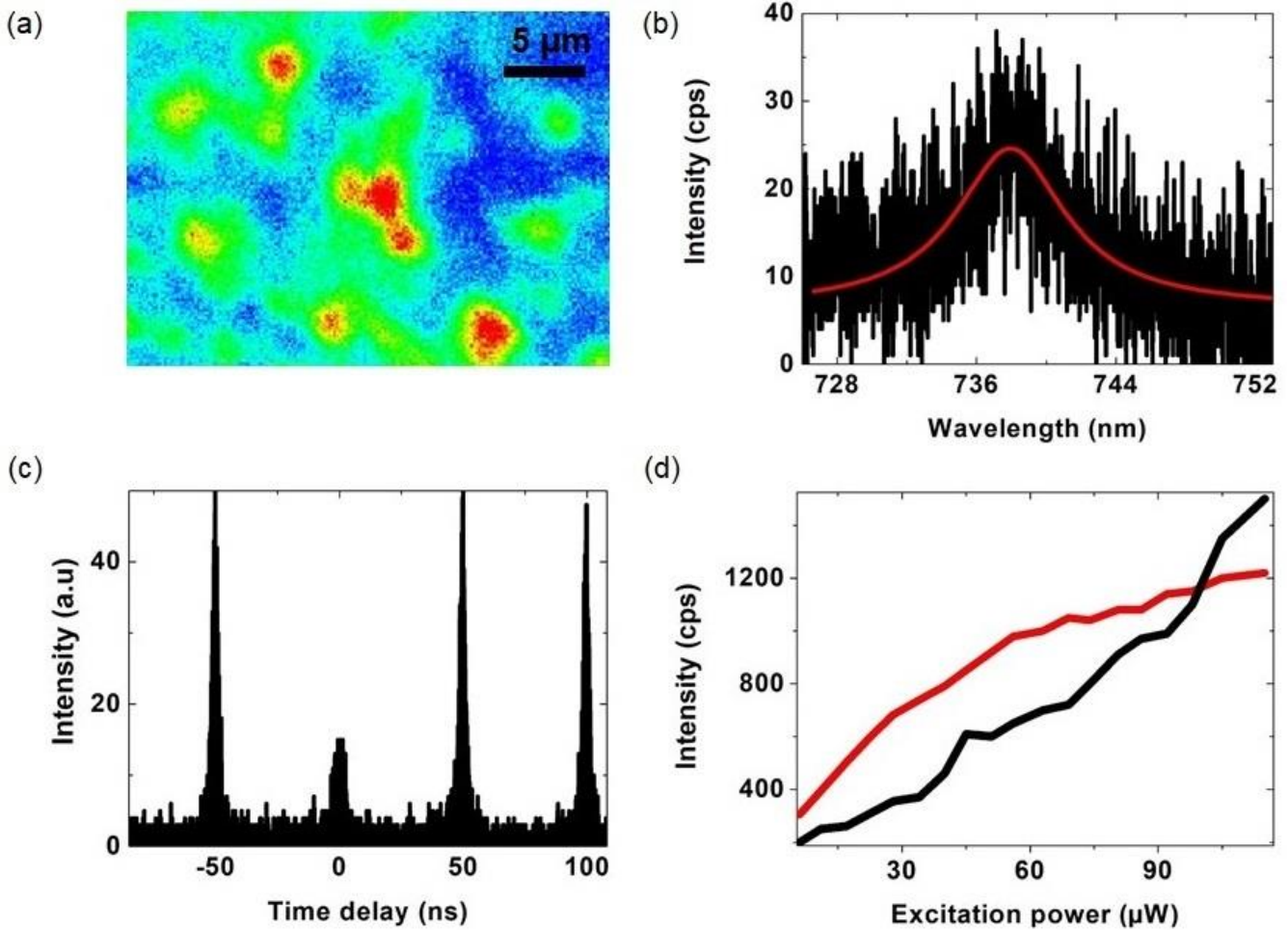

Figure 8. (a) Wide-field fluorescence image of SiV color centers in sample C. (b) Spectrum of a single SiV color center, where the red curve represents a Gaussian fit with a FWHM of 7.6 nm. (c) Photon anti-bunching under pulsed excitation. (c) Saturation (red curve) vs background (black curve) as a function of excitation power.

4. Conclusions

In conclusion, we have reported the creation and optical characterization of SiV color centers in P-doped diamond samples and have shown that the fluorescence background due to doping, nitrogen-impurities and ion-implantation induced defects can be significantly suppressed for attaining single-photon emission with high spectral quality. The most critical parameter is the nitrogen concentration in P-doped diamond, which should be below 1 ppb in the technical gases during CVD growth. The possibility of accessing individual SiV color centers in P-doped diamond paves the way to the implementation of simpler configurations for the electrical excitation of diamond-based single-photon sources, e.g. Schottky diodes in place of p-i-n junctions, hence facilitating the integration of quantum photonic devices in diamond-based electronics.

Acknowledgements

The authors gratefully acknowledge financial support from the University of Siegen, the German Research Foundation (DFG) (INST 221/118-1 FUGG, 410405168), the Hasselt University Special Research Fund (BOF), the Research Foundation- Flanders (FWO), and the Methusalem NANO network. SSN is a Newton International Fellow of the Royal Society. The authors also acknowledge INFN-CHNet, the network of laboratories of the INFN for cultural heritage, for support and precious contributions in terms of instrumentation and personnel. A.M. Flatae and M. Agio would like to thank F. Tantussi, F. De Angelis for experimental support and D. Yu. Fedyanin for helpful discussions. This work is based upon networking from the COST Action MP 1403 "Nanoscale Quantum Optics," supported by COST (European Cooperation in Science and Technology).

References

1. T. D. Ladd, F. Jelezko, R. Laflamme, Y. Nakamura, C. Monroe, J. L. O'Brien, Quantum computers, Nature 464, 45 (2010).
2. V. Giovannetti, S. Lloyd, L. Maccone, Advances in quantum metrology, Nat. Photonics 5, 222 (2011).
3. A. V. Sergienko, Quantum Communications and Cryptography (Boca Raton, FL: CRC, 2005)
4. Z. Yuan, B. E. Kardynal, R. M. Stevenson, A. J. Shields, C. J. Lobo, K. Cooper, N. S. Beattie, D. A. Ritchie, M. Pepper, Electrically driven single-photon source Science 295, 102 (2002).
5. D. J. P. Ellis, A. J. Bennett, S. J. Dewhurst, C. A. Nicoll, D. A. Ritchie, A. J. Shields, Cavity-enhanced radiative emission rate in a single photon-emitting diode operating at 0.5 GHz, New J. Phys. 10, 043035 (2008).
6. M. J. Conterio, N. Sköld, D. J. P. Ellis, I. Farrer, D. A. Ritchie, A. J. Shields, A quantum dot single photon source driven by resonant electrical injection, Appl. Phys. Lett. 103, 162108 (2013).
7. T. Heindel, C. Schneider, M. Lermer, S. H. Kwon, T. Braun, S. Reitzenstein, S. Höfling, M. Kamp, A. Forchel, Electrically driven quantum dot-micropillar single photon source with 34% overall efficiency, Appl. Phys. Lett. 96, 011107 (2010).
8. S. Lagomarsino, A.M. Flatae, S. Sciortino, F. Gorelli, M. Santoro, F. Tantussi, F. De Angelis, N. Gelli, F. Taccetti, L. Giuntini, M. Agio, Optical properties of silicon-vacancy color centers in diamond created by ion implantation and post-annealing, Diam. Relat. Mater. 84, 196 (2018).
9. D. Yu Fedyanin and M. Agio, Ultrabright single-photon source on diamond with electrical pumping at room and high temperatures, New J. Phys. 18, 073012 (2016).
10. Berhane A.M., Choi S., Kato H. Makino T., Mizuochi N., Yamasaki S. and Aharonovich I., Electrical excitation of silicon-vacancy centers in single crystal diamond, Appl. Phys. Lett. 106, 171102 (2015).
11. Tegetmeyer, B., Schreyvogel, C., Lang, N., Müller-Sebert, W., Brink, D. and Nebel, C.E., Electroluminescence from silicon-vacancy centers in diamond p-i-n diodes, Diam. Relat. Mater. 65, 42-46 (2016).
12. S. Koizumi, M. Kamo, Y. Sato, H. Ozaki, and T. Inuzuka, Growth and characterization of phosphorus doped {111} homoepitaxial diamond thin films, Appl. Phys. Lett. **71**, 1065 (1997).

13. S. Koizumi, T. Teraji, and H. Kanda, Phosphorus-doped chemical vapour deposition of diamond, Diamond Relat. Mater. **9**, 935 (2000).
14. T. A. Grotjohn, D. T. Tran, M. K. Yaran, S. N. Demlow, T. Schuelke, Heavy phosphorus doping by epitaxial growth on the (111) diamond surface, Diamond Relat. Mater. 44, 129 (2014).
15. Y. Balasubramaniam, P. Pobedinskas, S. D. Janssens, G. Sakr, F. Jomard, S. Turner, Y. –G. Lu, W. Dexters, A. Soltani, J. Verbeeck; j: Borjon; M. Nesladek, K. Haenen, Thick homoepitaxial (110)-oriented phosphorus doped n-type diamond, Appl. Phys. Lett. 109, 062105 (2016).
16. S. Bohr, R. Haubner, B. Lux, Influence of phosphorus addition on diamond CVD, Diam. Relat. Mater. 4, 113 (1995).
17. N. Orita, T. Nishimatsu, H. K. Yoshida, Ab initio study for site symmetry of phosphorus-doped diamond, Japan. J. Appl. Phys. 46, 315 (2007).
18. J. P. Goss, P. R. Briddon, R. Jones, S. Sque, Donor and acceptor states in diamond, Diam. Relat. Mater. 13, 684 (2004).
19. S. J. Sque, R. Jones, J. P. Goss, P. R. Briddon, Shallow donors in diamond: Chalcogens, Pnictogens, and their hydrogen complexes, Phys. Rev. Lett. 92, 017402 (2004).
20. J. Isoya, M. Katagiri, T. Umeda, S. Koizumi, H. Kanda, N. T. Son, A. Henry, A. Gali, E. Janzen, Pulsed EPR studies of phosphorus shallow donors in diamond and SiC, Physica, 376, 358 (2006).
21. H. Kato, T. Makino, S. Yamasaki, H. Okushi, n-type diamond growth by phosphorus doping on (001)-oriented surface, J. Phys. D: Appl. Phys. 40, 6189 (2007).
22. J. Barjon, Luminescence spectroscopy of bound excitons in diamond, Phys. Status Solidi A, 214, 1700402 (2017).
23. J. te Nijenhuis, S. M. Olsthoorn, W. J. P. van Enckevort, L. J. Giling, Red luminescence in phosphorous-doped chemically vapor deposited diamond, J. Appl. Phys. 82, 419 (1997).
24. R. Jones and J. E. Lowther, J. Goss, Limitations to n-type doping in diamond: The phosphorus-vacancy complex, Appl. Phys. Lett. 69, 2489 (1996).
25. G. Z. Cao, F. A. J. M. Driessen, G. J. Bauhuis, L. J. Giling, P. F. A. Alkemade, Homoepitaxial diamond films codoped with phosphorus and nitrogen by chemical-vapor deposition, J. Appl. Phys. 78, 3125 (1995).
26. S. Dannefaer, W. Zhu, T. Bretagnon, and D. Kerr, Vacancies in polycrystalline diamond films, Phys. Rev. B **53**, 1979 (1996).
27. J.F. Prins, Activation of boron-dopant atoms in ion-implanted diamonds, Phys. Rev. B 38, 5576–5584 (1988).
28. R. Kalish, Ion implantation in diamond for quantum information processing (QIP): doping and damaging, Quantum Information Processing with Diamond, Principles and Applications, Woodhead Publishing, pp. 36–67 (2014).
29. S. Lagomarsino, S. Sciortino, N. Gelli, A.M. Flatae, F. Gorelli, M. Santoro, M. Chiari, C. Czelusniac, M. Massi, F. Taccetti, M. Agio, L. Giuntini, The center for production of single-photon emitters at the electrostatic deflector line of the Tandem accelerator of LABEC (Florence), Nucl. Instr. Meth. Phys. Res. B 422, 31 (2018).

30. A. Tallaire, T. Ouisse, A. Lantreibecq, R. Cours, M. Legros, H. Bensalah, J. Barjon, V. Mille, O. Brinza, J. Achard, Identification of dislocations in synthetic chemically vapour deposited diamond single crystals, Cryst. Growt Des. 16, 2741 (2016).